# Highly Conducting Graphene Sheets and Langmuir-Blodgett Films

Xiaolin Li[†], Guangyu Zhang[†], Xuedong Bai[‡], Xiaoming Sun[†], Xinran Wang[†], Enge Wang[‡], Hongjie Dai[†]*

[†] *Department of Chemistry and Laboratory for Advanced Materials, Stanford University, Stanford, CA 94305, USA*

[‡] *Institute of Physics, Chinese Academy of Sciences, Beijing 100080, China*

* Correspondence to hdai@stanford.edu

Graphene is an intriguing material with properties that are distinct from those of other graphitic systems.[1-5] The first samples of pristine graphene were obtained by 'peeling off' [2, 6] and epitaxial growth [5, 7]. Recently, the chemical reduction of graphite oxide was used to produce covalently functionalized single-layer graphene oxide. [8-15] However, chemical approaches for the large-scale production of highly conducting graphene sheets remain elusive. Here, we report that the exfoliation-reintercalation-expansion of graphite can produce high-quality single-layer graphene sheets stably suspended in organic solvents. The graphene sheets exhibit high electrical conductance at room and cryogenic temperatures. Large amounts of graphene sheets in organic solvents are made into large transparent conducting films by Langmuir-Blodgett (LB) assembly in a layer-by-layer manner. The chemically derived high quality graphene sheets could lead to future scalable graphene devices.

Several methods have been explored thus far to obtain graphene in solution phase via chemical routes. Graphite oxide (GO) was prepared by harsh oxidation using the Hummer's method.[16] The as-made GO was electrically insulating but chemical reduction[9,10,15] partially recovered the conductivity, albeit at values orders of magnitude below that of pristine graphene. Irreversible defects and disorder exist in the GO sheets.[9,10] The reduced GO exhibit non-metallic behavior, with the conductance decreasing by about three orders of magnitude upon cooling to low temperature,[13] whereas pristine graphene is nearly metallic.[2,17] Recently, we obtained pristine graphene nanoribbons (GNR) by sonicating thermally exfoliated graphite in a 1,2-dichloroethane (DCE) solution of poly(m-phenylenevinylene-co-2,5-dioctoxy-p-phenylenevinylene)(PmPV).[18]
Nevertheless, the yield was low and most of the ribbons contained two or more layers. Despite these and other efforts,[8-15,19-22] solution phase derivation of single-layer graphene with high electrical conductivity from widely available parent graphite materials has not been achieved at a large scale. The production of stable suspensions of graphene in organic solvents is also an important goal in chemical processing and other areas.

In the current work, to make high quality graphene sheet (GS), we started by first exfoliating commercial expandable-graphite (160-50N, Grafguard Inc.) by brief (60s) heating to 1000°C in forming gas. We then ground the exfoliated graphite, re-intercalate with oleum (fuming sulfuric acid with 20% free $SO_3$), and inserted tetrabutylammonium hydroxide (TBA, 40% solution in water) into oleum intercalated graphite (Fig. 1a) in N, N-dimethylformamide (DMF, see Method). We then sonicated the TBA-inserted oleum-intercalated graphite (Fig. 1b) in a DMF solution of 1,2-distearoyl-sn-glycero-3-phosphoethanolamine-N-[methoxy(polyethyleneglycol)-5000] (DSPE-mPEG) for 60 mins to form a homogeneous suspension. Centrifugation was used to remove large pieces of materials from the supernatant (Fig. 1c) (See Methods). This method easily obtained large amounts of graphene sheets suspended in DMF and could be transferred to other solvents including water and organic solvents.

We used atomic force microscopy (AFM) to characterize the materials deposited on substrates from the supernatant and observed ~90% single layer GS with various shapes and sizes (Fig. 1d). For over hundreds of graphene sheets measured, we found that the single-layer GS have an average size of about 250nm (Fig. S2a) and topographic height of ~ 1nm (Fig. S2b and Fig. S3). Transmission electron microscopy (TEM, Fig. 1e) and electron diffraction (ED, Fig. 1f) were used to characterize the single layer GS. The ED pattern of our GS was similar to that of 'peeled off' graphene,[23] suggesting well-crystallized single layer graphene structure.

Our starting expandable graphite was prepared by chemical intercalation of sulfuric acid and nitric acid.[24] Upon heating, they exfoliated violently due to volatile gaseous species released from the intercalant. Most of the exfoliated graphite was still in multi-layer graphene form.[25] In order to get single layer graphene sheets, we invoked re-intercalation by oleum, a chemical known to strongly debundle carbon nanotubes due to intercalation.[26] TBA was a molecule capable of inserting and expanding the distance between heavily oxidized graphite layers.[27] We suggest that TBA also insert into oleum-intercalated graphite to increase the distance between adjacent graphitic layers (Fig.1a), which facilitated the separation of graphene sheets upon sonication in a surfactant solution.[27] This was evidenced by that without the TBA treatment step, the yield of single layer GS was extremely low by the otherwise identical method (see supplementary Fig. S1 for control experiments). We also found that DMF was a better solvent than water for our method. Further, DSPE-mPEG was a surfactant capable of suspending nanotubes,[28] and was another important factor to obtaining homogeneous suspension of GS.

Our weak oleum treatment condition (soaking in oleum at room temperature for one day) is important to obtain high quality GS without excessive chemical functionalization and thus property degradation. The conjugate graphene plane is largely free of irreversible modifications through the treatment steps. Room temperature oleum treatment is much less oxidative than the Hummer's method, evidenced by the as-made GS exhibiting significantly fewer functional groups (Fig. 2a and b) than as-made Hummer's GO (Fig. 2d and e) in infrared (IR) spectra. The IR spectrum of as-made GS (Fig. 2a) showed weaker signals of carboxylic groups than the Hummer's GO (shading range in Fig. 2a and 2d) [29]. X-ray photoelectron spectroscopy (XPS) (Fig. 2b) of our as-made GS showed small but noticeable signals at higher binding energy corresponding to small amount of C-O species.[9, 29] These species were removed by 800°C $H_2$ annealing, indicating the formation of high quality graphene (Fig. 2b). The annealed GS exhibited the same XPS spectrum as a pristine highly oriented pyrolytic graphite (HOPG) crystal (Fig. 2b), confirming the lack of significant defects or covalently modifications of sp2 carbon in the final GS product.

We propose the schematic structures of the intermediate and final product of our GS and Hummer's GO in Fig. 2c and f. Oxidization of our intermediate, as-made GS was relatively mild and the few covalently attached functional groups such as carboxylic group (seen in the IR spectrum Fig. 2a) and hydroxyl group were most likely at the edges of as-made GS (Fig. 2c). This was supported by the fact that our as-made GS showed similar electrical conductivity as 800°C vacuum-annealed GS (see Fig. 3b and c), an unlikely result if the graphene plane was heavily modified covalently. The Hummer's GO was heavily oxidized with disrupted conjugation in the plane, missing carbon atoms in the plane,[30] and abundant functional groups such as epoxide, hydroxyl, carbonyl and carboxyl groups at both the edges and in the plane (Fig. 2f).[9, 10] Importantly, these abundant functional groups weaken the van der Waals interactions between the layers of GO and make them hydrophilic, which is the reason of single-layer GO exfoliation in aqueous media to form stable suspensions without the need of insertion agent such as TBA or the assistance of surfactant for suspension. Thermal annealing removed some of the functional groups but was unable to completely repair the holes and other irreversible defects formed within the plane of Hummer's GO sheets (Fig. 2f).[9, 10]

We fabricated single GS electrical devices with as-made and annealed GS and Hummer's GO. We used palladium (Pd) or titanium/gold (Ti/Au) as source/drain (S/D) metal contacts (channel length L~100nm), a $P^{++}$-Si backgate, and 500nm $SiO_2$ as gate dielectrics (Methods and supplementary information). Typical resistance of ~100 nm wide GS (Fig. 3a) at room temperature is 10-30 kOhm (Fig. 3b and 3c). The average resistance histogram (error bar is the standard deviation) for large numbers of devices showed that room-temperature resistance of as-made GS was similar to those of annealed GS devices (for both Pd and Ti/Au contacted devices), and about 100 times lower than annealed GO (Fig. 3b). As-made GO devices without annealing were all electrically insulating. This result strongly supported the proposed atomic structures of GS and GO (Fig. 2c and 2f) and that our GS are nearly pristine graphene. Our thermally annealed GS retained high electrical conductivity with only slight increase in resistance at low temperatures (for both Pd and Ti/Au contacted devices), in strong contrast to annealed GO that were insulating at low temperatures (Fig. 3c). Devices of as-made GS showed reduced metallic characteristics over annealed GS devices (but were still >1000 times more conducting than GO devices) with larger increase in resistance at low T (Fig. 3c). This suggested that the as-made GS contained a small amount of disorder in the structures.

To explore the utility of our high quality graphene sheet, we transferred large quantities of GS from DMF to organic solvent DCE with excellent stability against agglomeration. The fact that our as-made GS was stably suspended in DCE without additional surfactant indicates high hydrophobicity of the graphene, consistent with low degree of graphene oxidation and covalent functionalization. In contrast, Hummer's GO were highly hydrophilic and completely insoluble in organic solvents. The organic stability of our GS enabled Langmuir-Blodgett (LB) films to be made on various transparent substrates including glass and quartz (see Methods and supplementary information) for producing transparent and conducting films. This was done by adding GS suspensions onto water subphase, vaporizing the DCE solvent from water surface, compressing the floating GS and transferring the GS LB film onto a substrate by dip-coating. The GS floated on water due to hydrophobicity within the sheet. The edges of GS contain functional groups, giving rise to planar amphiphilic species. We were able to transfer GS repeatedly to achieve multi-layer films. The 1-, 2-, and 3-layer LB films on quartz (Fig. 4a) afforded a sheet resistance of ~150k, 20k, and 8k ohm at room temperature (Fig. 4c) and a transparency (defined as transmittance at 1000nm wavelength) of ~93%, 88% and 83% respectfully (Fig.4b and 4c). With 3-layer LB film, we can easily reach 8 kOhm sheet resistance with the transparency higher than 80%, which compares favorably over reduced GO films.[11, 12] The conductance and transparency of our films are comparable to those made of graphene sheets formed by sonication of natural graphite in dimethylformamide.[31] This is the first time that high quality graphene sheets were assembled by the LB technique in a layer-by-layer manner on large substrates. Note that with the same method, we also succeeded in making GS using pristine graphite flakes as the starting material, and the structural, electrical and spectroscopic properties of the GS made from pristine flakes are similar to those made from expandable graphite. Thus, our large-scale synthesis of graphene sheet and the ability of processing them in various solvents for assembly open up the door to high-performance, scalable applications such as solar cells using transparent conducting films.

**Methods**

**Preparation of GS suspension**

Our single layer graphene sheets (GS) preparation started by exfoliating expandable graphite (160-50N of Grafguard Inc.) at 1000°C in forming gas for 60s. Then exfoliated graphite (~10mg) was ground with NaCl crystallites for 3 mins forming a uniform grayish mixture. Small pieces of exfoliated graphite were separated and collected by dissolving NaCl with water and filtration. The resulting sample was then treated with oleum at room temperature for a day. After complete removal of acid by filtration and repeated washing, the resulting sample was ultra-sonicated using a cup-horn sonicator in DMF (10mL) solution of TBA (130μl) for 5mins. The suspension was put at room temperature for 3 days to let the TBA fully inserted into graphene layers. Then 5mL suspension was taken out and bath-sonicated with DSPE-mPEG (Laysan Bio. Inc., Arab, Alabama) (15mg) for 1hr forming a homogeneous suspension. After centrifuging the suspension at 24kg for 3mins, we obtained black suspension with mostly single layer GS retained in the supernatant.

**Characterization of GS**

AFM images of GS were taken with a Nanoscope IIIa multimode instrument. The samples were prepared by soaking a SiO$_2$ substrate (pretreated by 4mM 3-aminopropyltriethoxysilane (APTES) water solution for 20 mins) into the graphene suspension for 20 mins, rinsing with water and blow-dry with Argon. The substrate was calcined to 350°C and annealed at 800°C in $H_2$ before AFM. IR spectrum (400 to 4000cm$^{-1}$) was measured using Nicolet IR100 FT-IR Spectrometer with pure KBr as the background. After removal of the surfactant by filtration and repeated washing, graphene sample was collected and ground with KBr. The mixture was dried and compressed into a transparent tablet for measurement. We characterized our GS by a JEOL 2010F FEG transmission electron microscope (TEM) at an accelerating voltage of 120kV. The TEM samples were prepared by drying a droplet of the graphene suspension on a lacey carbon grid. High resolution XPS measurement was carried out using SSI S-Probe Monochromatized XPS Spectrometer, which uses Al (Kα) radiation as a probe. Analysis spot size is 150 micron by 800 micron. Sample preparation involved removal of the surfactant by filtration and repeated washing, depositing materials onto a silicon substrate by repeated drop-drying. GO sample was prepared by depositing materials onto a silicon substrate by repeated drop-drying. HOPG sample was used for XPS measurement without any treatment.

**GS and GO device fabrication**

GS and GO were deposited onto 500nm $SiO_2$/P$^{++}$Si substrate (Pre-treated with 4mM APTES solution). After removal of the surfactant by 350°C calcination and 800°C $H_2$ annealing, we used electron-beam lithographic patterning followed by electron-beam evaporation of Pd (20nm) or Ti(1.5nm)/Au(20nm) to form source and drain electrodes (channel length ~100nm, width ~2micron) on the substrate randomly. The sample was then annealed in argon at 300˚C for 15 min to improve the contacts between the source and drain metal and the GS/GO in the channel region.

**LB film fabrication**

DMF suspension of GS was centrifuged at 24kg for 1hr to remove the surfactants. The aggregates were then re-suspended in fresh DMF by brief sonication. This centrifugation and re-suspending process was repeated for 3 times. Then we re-suspend the GS samples in fresh DCE and repeat the centrifugation and re-suspending process for 3 times to ensure complete removal of DSPE-mPEG. The resulting GS were suspended in DCE by 5mins sonication. GS LB films were made using a commercial KSV-Minimicro 2000 LB trough. About 1.2mL of GS/DCE suspension was added to a water subphase in the LB trough. A platinum plate was used to monitor the surface tension during compression of the GS on the water subphase by moving the two opposing barriers towards each other. At a target surface pressure of ~27mN/m, GS were compressed to form a dense LB film transferable onto a solid substrate (up to 1x1 in$^2$) by slowly pulling up the substrate out of the aqueous subphase. The transferred GS LB film was typically calcined at 350 ºC to remove DSPE-mPEG and TBA residues before transparency and resistance measurement. After calcination the quartz substrate with 1 layer LB film we then transfer another layer GS film onto it by repeating the LB making procedure. We are able to obtain multi-layer LB films by this layer-by-layer transfer method. Transparency of the GS films was measured with Cary 6000i spectrophotometer using pure quartz as the background. The transparency was defined as the transmittance at 1000nm wavelength.

**FIGURE CAPTIONS**

**Figure 1.** Chemically derived single layer graphene sheets (GS) from solution phase. **(a)** Schematic drawing of the exfoliated graphite re-intercalated with sulphuric acid molecules (teal spheres) between the layers. **(b)** Schematic drawing of TBA (blue spheres) insertion into the intercalated graphite. **(c)** Schematic drawing of GS coated with two DSPE-mPEG molecules and a photograph of a DSPE-mPEG/DMF solution of GS. **(d)** AFM image of typical GS with size of about several hundred nanometers and topographic height of about 1nm. (See supplementary information for height details.) Scale bar is 300nm. **(e)** Low magnification TEM images of a typical GS with the size of about several hundred nanometers. Scale bar is 100nm. **(f)** Electron diffraction (ED) pattern of an as-made GS in **(e)** showing excellent crystallization of the GS.

**Figure 2.** Graphene sheets versus graphene oxide sheets. **(a)** IR spectrum (400-4000$cm^{-1}$) of as-made GS. The shading region is from about 1400 to 1900$cm^{-1}$ showing the signal of carboxylic groups. **(b)** XPS spectra of as-made, annealed GS and a HOPG crystal. Note the similarity between the spectra of the annealed GS and HOPG. **(c)** Schematic drawing of the atomic structure of as-made and annealed GS. **(d)** IR spectrum (400-4000$cm^{-1}$) of as-made GO. The shading region is from about 1400 to 1900$cm^{-1}$ showing the signal of carboxylic groups. **(e)** XPS spectra of as-made, annealed GO and a HOPG crystal. **(f)** Schematic drawing of the atomic structure of as-made and annealed GO.

**Figure 3.** Electrical characterization of single GS. **(a)** AFM image of a typical device with single graphene sheet (GS, thickness~1nm, single layer) bridging the channel (channel length $L$ ~ 100nm GS with TiAu contacts and Si back-gate) between source (S) and drain (D) electrodes. Scale bar is 200nm. **(b)** Mean resistance histogram of 10 devices of as-made GS, annealed GS and annealed GO. The resistance of as-made GS and annealed GS were similar (within the error bars of statistical variations between GS

devices) indicating the high quality of our as-made GS. **(c)** Resistance of as-made GS, annealed GS and annealed GO at various temperatures. The resistance of GS, especially the annealed GS showed only very small conductance drop (similar to some of the peel-off pristine graphene samples reported in the literature) at low temperature. Black curve is the resistance of 800ºC annealed GS (Ti/Au contact); red curve is the resistance of 800ºC annealed GO; green curve is the resistance of as-made GS; blue curve is the resistance of 800ºC annealed GS (Pd contact).

**Figure 4.** Large-scale LB films of GS. **(a)** A photograph of 2-layer GS LB film with part of it left clear. Scale bar is 10mm. **(b)** Transparency spectra of 1- (black curve), 2- (red curve), and 3- (green curve) layer GS LB film. The transparency was defined as the transmittance at 1000nm wavelength. **(c)** Resistance and transparency of 1-, 2-, and 3-layer LB film. The small percentage of bi-layer and few-layer GS in our sample and GS overlapping in the LB film over the substrate contributed to the transparency loss.

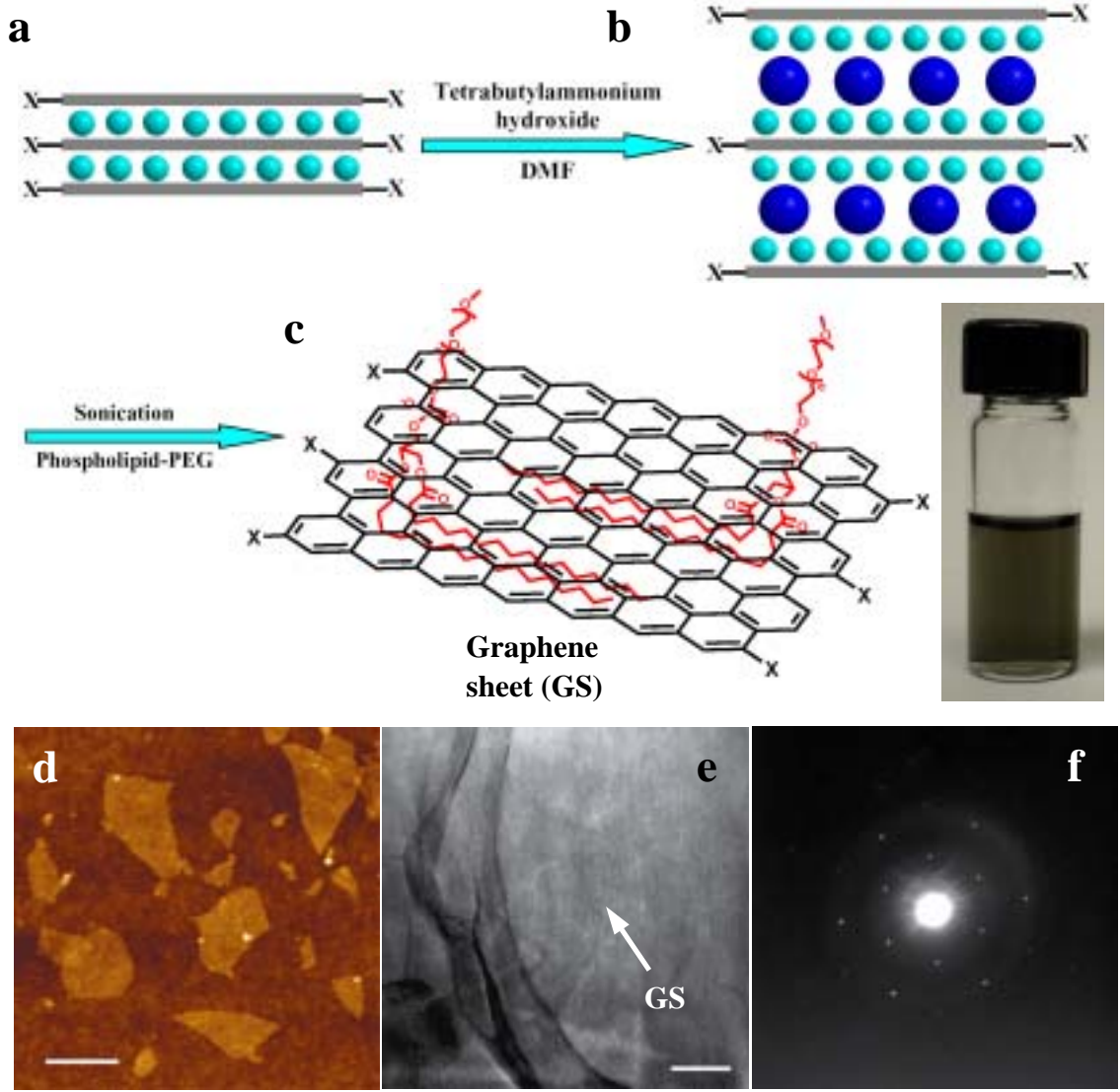

Figure 1

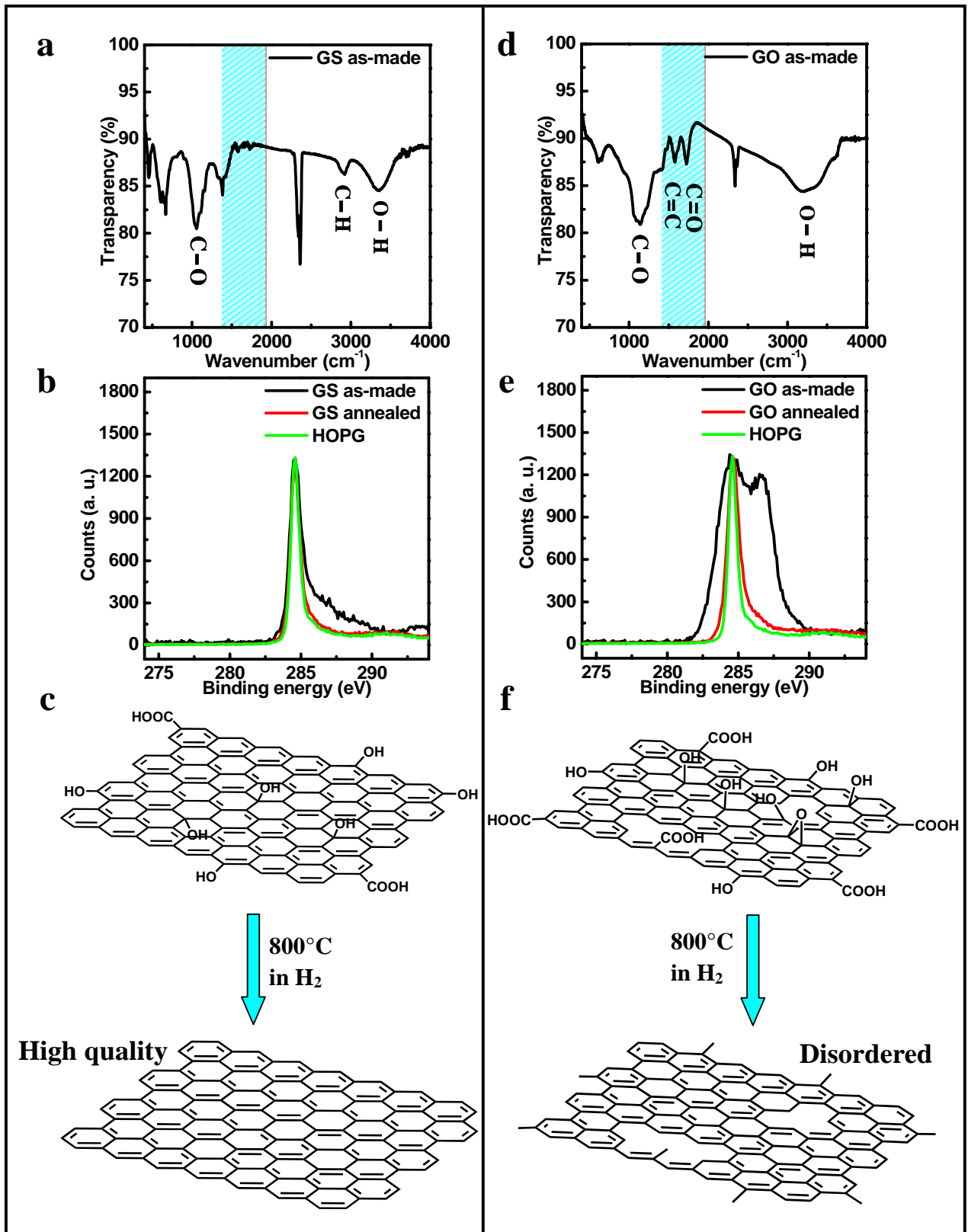

Figure 2

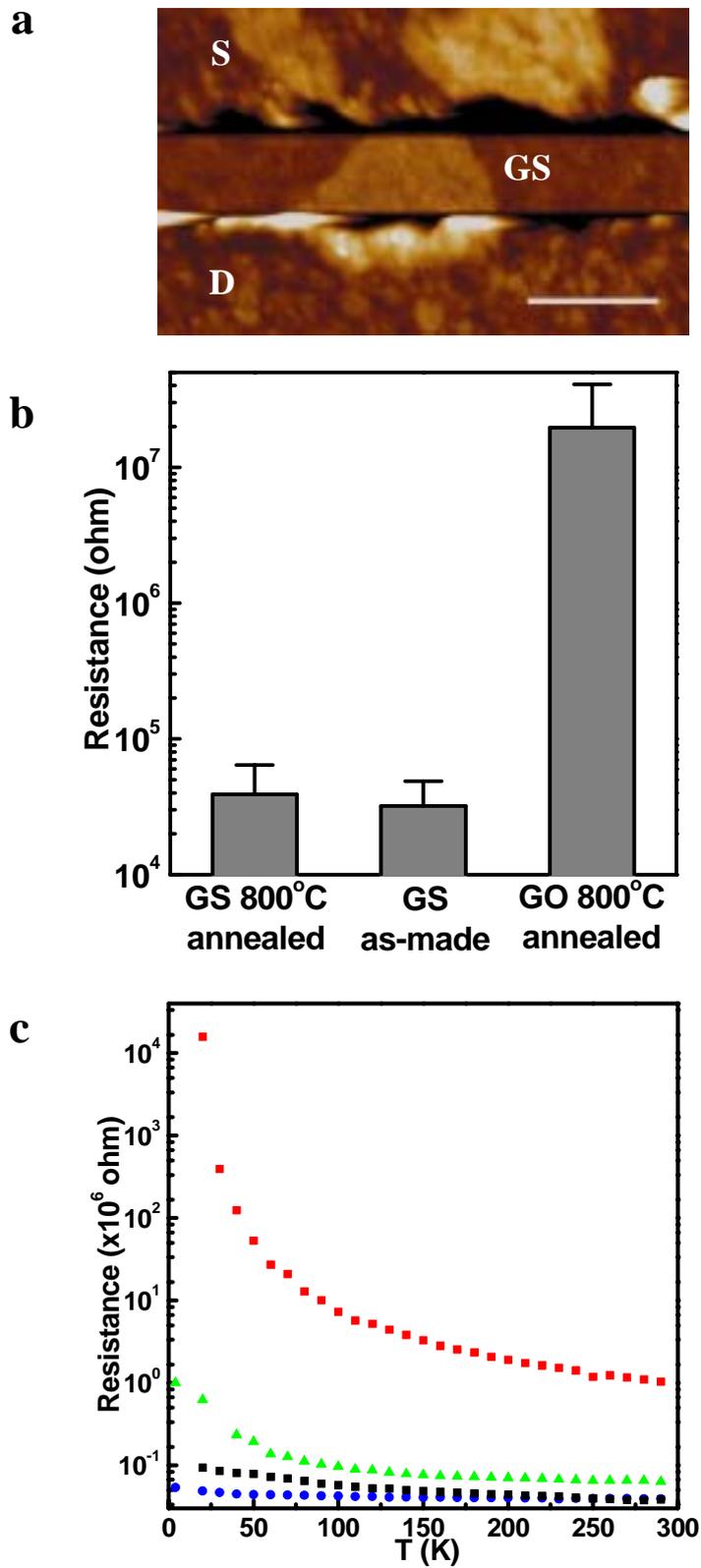

Figure 3

**a**

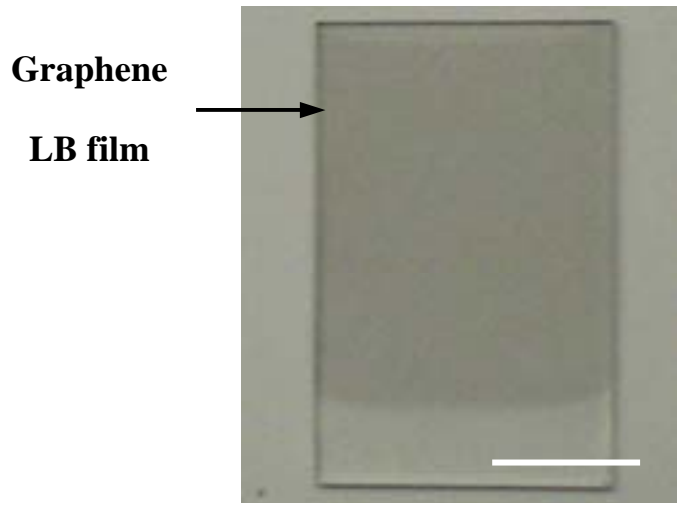

Graphene LB film →

**b**

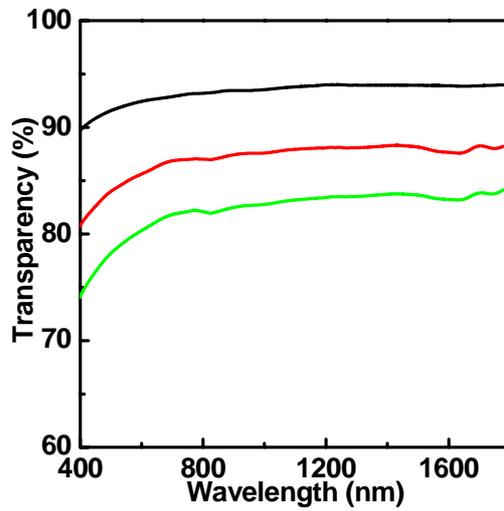

**c**

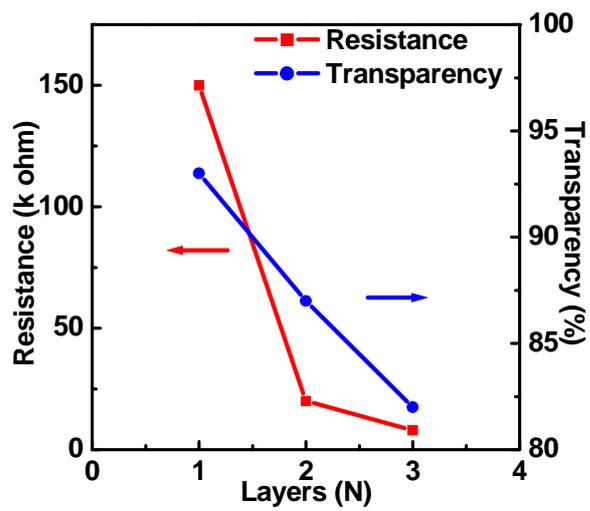

Figure 4